\documentclass[aps,prl,showpacs,notitlepage,nofootinbib,superscriptaddress,floatfix,showkeys,twocolumn]{revtex4-1}

\usepackage{blindtext}
\usepackage{hyperref}
\usepackage{amsmath,amssymb}
\usepackage{float}
\usepackage{microtype}
\usepackage{graphicx}
\usepackage{bm}
\usepackage{latexsym}
\usepackage{epsfig}
\usepackage{psfrag}
\usepackage{color}
\usepackage[dvipsnames]{xcolor}
\usepackage{subfigure}
\usepackage{multirow}
\usepackage{colortbl}
\usepackage[section]{placeins}
\def\e1i{\epsilon_{1\mathrm{i}}}

\allowdisplaybreaks[1]

\begin{document}

\title{Robust constraints on feebly interacting particles using XMM-Newton}

\author{Pedro De la Torre Luque}\email{pedro.delatorreluque@fysik.su.se}
\affiliation{The Oskar Klein Centre, Department of Physics, Stockholm University, Stockholm 106 91, Sweden}

\author{Shyam Balaji}
\email{sbalaji@lpthe.jussieu.fr}
\affiliation{Laboratoire de Physique Th\'{e}orique et Hautes Energies (LPTHE),
UMR 7589 CNRS \& Sorbonne Universit\'{e}, 4 Place Jussieu, F-75252, Paris, France}
\affiliation{Institut d'Astrophysique de Paris, UMR 7095 CNRS \& Sorbonne Universit\'{e}, 98 bis boulevard Arago, F-75014 Paris, France}

\author{Pierluca Carenza}\email{pierluca.carenza@fysik.su.se}
\affiliation{The Oskar Klein Centre, Department of Physics, Stockholm University, Stockholm 106 91, Sweden}
\smallskip

\begin{abstract}
During galactic Supernova (SN) explosions, a large amount of feebly interacting particles (FIPs) may be produced. In this work we analyze electrophilic FIPs with masses in the MeV-range that escape from SN and decay into electron-positron pairs, causing an exotic leptonic injection. This contribution adds up to known components, leading to an unexpected excess of X-ray fluxes generated by inverse-Compton scattering of the injected particles on low-energy photon backgrounds. For the first time in the context of FIPs, we use XMM-Newton X-ray measurements to obtain the strongest and most robust bounds on electrophilic FIPs produced by SN in our Galaxy.
\end{abstract}
\maketitle

\emph{Electrophilic FIPs.}
Massive stars explode in Supernovae (SN), reaching nuclear densities and very high temperatures in the core. These extreme environmental conditions are ideal to abundantly produce light Feebly Interacting Particles (FIPs), if they exist. Currently, a lot of attention is devoted to FIP phenomenology from the point of view of theoretical and experimental investigation~\cite{Antel:2023hkf}. The large family of FIPs includes, among others, axions and axion-like particles (ALPs)~\cite{Raffelt:1987yt,Keil:1996ju,Carenza:2019pxu,Caputo:2021rux}, sterile neutrinos \cite{Kolb:1996pa,Raffelt:2011nc,Mastrototaro:2019vug}, light CP-even scalars \cite{Balaji:2022noj,Balaji:2023nbn} and dark photons (DPs) \cite{Chang:2016ntp}.  
Once produced in SN, because of their weak interaction with matter, FIPs may escape the stellar core and compete with neutrinos in subtracting energy to the SN. This in turn leads to a shortening of the neutrino burst signal~\cite{Raffelt:2012kt,Chang:2018rso}. 
The importance of SN for FIP phenomenology is not limited to indirect consequences on the neutrino burst, but also for direct signatures of FIP interactions and decays outside the star.

\begin{figure}[t!]
\begin{minipage}{\columnwidth}
\includegraphics[width=\columnwidth]{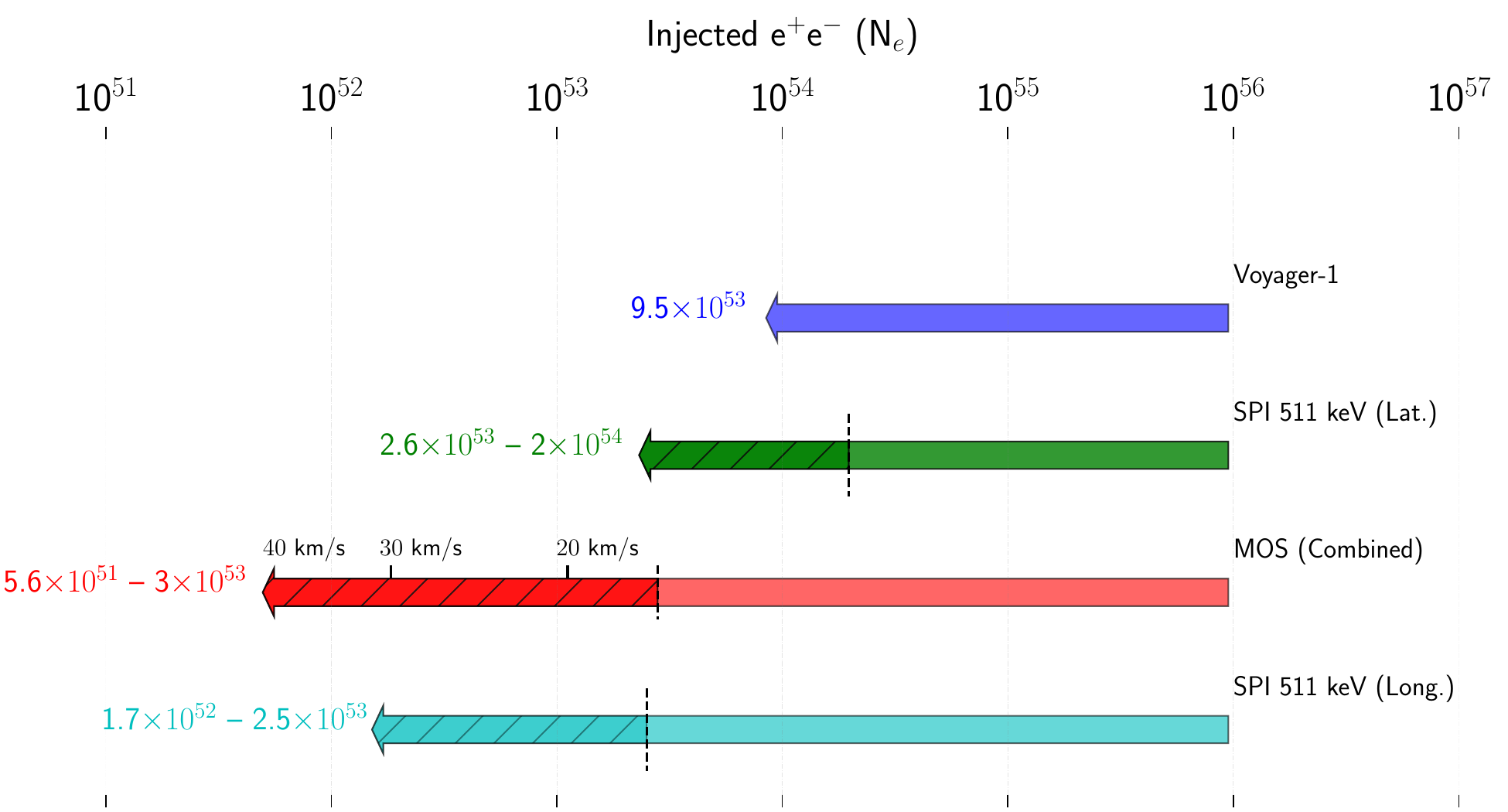}
\end{minipage} 
\caption{Limits on the production of electrons-positrons from FIP decays at $95\%$ Confidence Level (C.L.). We compare the limits derived from the fit to Voyager-1 data (blue), SPI latitude (green) and longitude (cyan) profiles, from Ref.~\cite{DelaTorreLuque:2023huu}, with the ones obtained in this work from observations of the MOS detector (red). The main systematic uncertainties in the evaluation of these limits are represented as hatched bars, which, in the case of MOS, correspond to uncertainties on electron/positron reacceleration  by plasma turbulence. 
These limits are valid for FIP masses below $\sim20$~MeV.}
\label{fig:Limits_Arrow}
\end{figure}

In this work we analyze FIPs with masses in the MeV-range (we focus on low FIP masses; below $\sim20$~MeV, where the injected spectrum barely changes), produced in SN, and decaying into electron-positron pairs outside the progenitor star. We refer to these particles as `electrophilic'~\cite{Appelquist:2002me, Asaka:2005an, Abel:2008ai, Jaeckel:2010ni, Alekhin:2015byh}. Several motivated FIPs belong to this class of exotic particles, such as ALPs, sterile neutrinos and DPs.
The exotic injection of electrons and positrons by this mechanism summed up with the standard channels, mainly pulsar wind nebulae, SN remnants~\cite{1969ocr..book.....G, Gabici:2019jvz} and cosmic ray (CR) interactions with the interstellar gas~\cite{DelaTorreLuque:2023zyd}, leads to important observational consequences that we will explore.
For instance, positrons would annihilate with background electrons and contribute to the monochromatic signal at 511~keV~\cite{Dar:1986wb}. Requiring that positrons injected by FIP decays do not produce a signal exceeding the one measured by the Spectrometer on INTEGRAL (SPI) instrument~\cite{Bouchet:2010dj,Siegert:2015knp}, bounds on various FIP models were determined~\cite{DeRocco:2019njg,Calore:2021klc,Calore:2021lih}.


We have explored, for the first time, a different phenomenology and probe of FIPs. We characterize the inverse-Compton (IC) emission generated by the injected leptons when scattering on the photon background. 
We use the X-ray observations from XMM-Newton in the range $2.5$-$8$~keV to place robust constraints on electrophilic FIPs, setting the stage for future studies of these data to probe FIPs. As we see from Fig.~\ref{fig:Limits_Arrow}, where we show the upper limits in the number of electrons/positrons produced in the Galaxy by SN-injected FIPs for different observables, XMM-Newton data (from the MOS detector, in red) may lead to the most stringent constraints up to date, even stronger than the emission line at $511$~keV (longitudinal profile, cyan).

\emph{FIP-induced electron-positron flux injection.} Electrophilic FIPs are defined as FIPs coupling mainly to electrons and positrons. Importantly for our discussion, when their mass exceeds $\sim1$~MeV the, dominant decay channel is into electron-positron pairs. Common examples are ALPs ($a$) decaying as $a\to e^{+}e^{-}$ or sterile neutrinos, where a possible decay channel is $\nu_{s}\to \nu_{\mu}e^{+}e^{-}$. In this second case, even though the FIP produces other particles in the decay, we can model it in a similar way as any electrophilic FIP. There are many other representative cases, like scalars, Kaluza-Klein gravitons and supersymmetric particles. Therefore, electrophilic FIPs are commonly found in Standard Model extensions and it motivates our discussion on their phenomenology.

\begin{figure*}[t!]
  \begin{minipage}{.45\textwidth}
    \includegraphics[width=\linewidth]{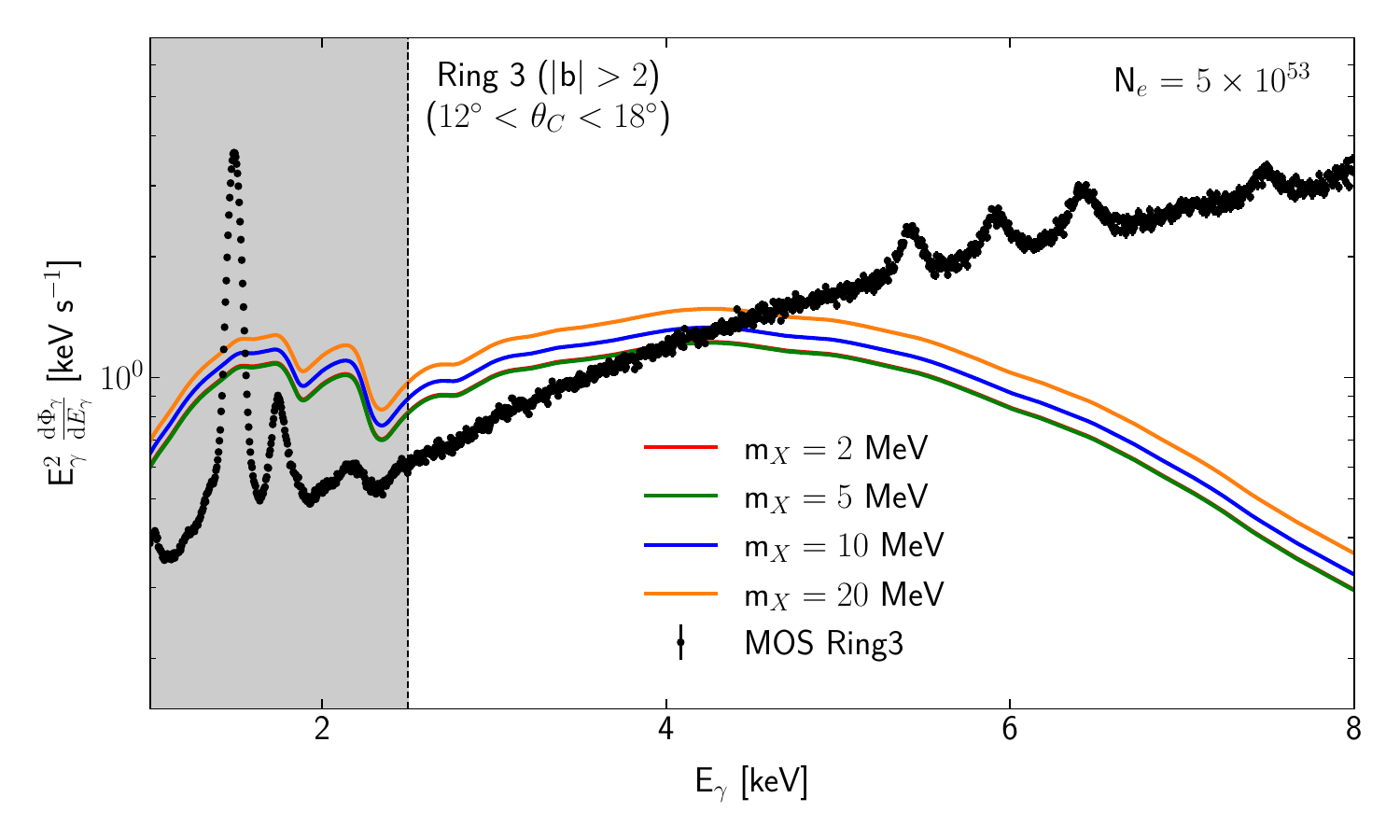}
  \end{minipage} \quad
  \begin{minipage}{.45\textwidth}
    \includegraphics[width=\linewidth]{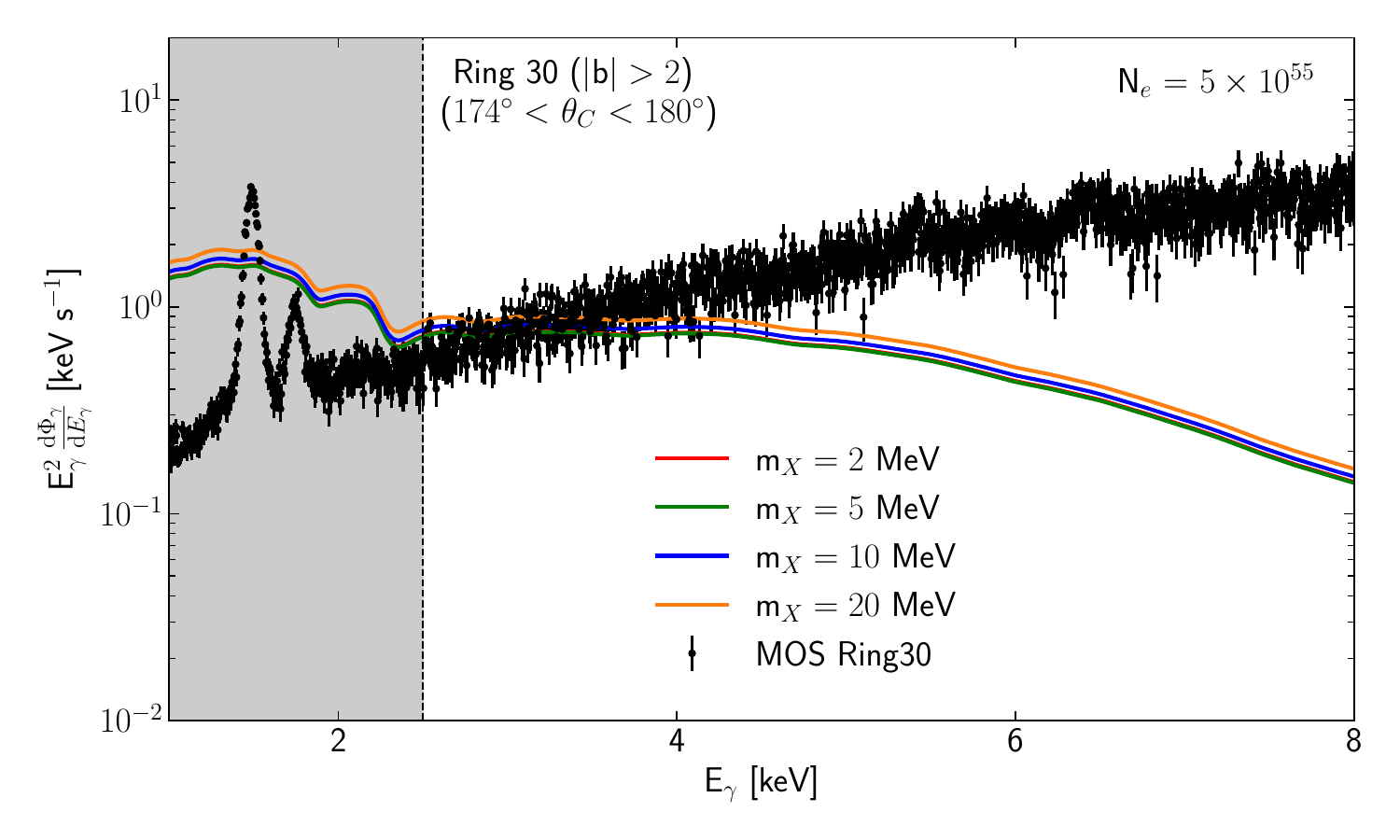} 
  \end{minipage} \quad
\caption{Comparison of the predicted X-ray signals generated from electrophilic FIPs from $1$ to $8$~keV, compared to the data from the MOS detector (XMM-Newton) for the rings 3 (the most constraining), in the left panel, and ring 30 (the less constraining), in the right panel. We show the predicted FIP X-ray emission for FIP masses ranging from $2$ to $20$~MeV, where FIPs produce roughly the same emission. The region below $2.5$~keV, not considered in our analysis, is shaded.}
\label{fig:XMM-Masses}
\end{figure*}

After being generated during galactic SN explosions, FIPs escape the parent star and undergo a decay process that populates the diffuse electron-positron Galactic background. The spatial distribution of this injection roughly follows the SN distribution, being affected by propagation effects.
The electron/positron flux caused by FIPs has an energy distribution inherited by the parent FIP. Assuming that electrons/positrons have an energy $E_{e}$ which is half of the decaying FIP, the injected flux can be parametrized as \cite{vanBibber:1988ge}
\begin{equation}
\begin{split}
    \frac{dN_{e}}{dE_{e}}&=N_{e}C_{0}\left(\frac{4E_{e}^{2}-m_{X}^{2}}{E_{0}^{2}}\right)^{\beta/2}e^{-(1+\beta)\frac{2E_{e}}{E_{0}}}\,,\\
      C_{0}&=\frac{2\sqrt{\pi}\left(\frac{1+\beta}{2m_{X}}\right)^{\frac{1+\beta}{2}}E_{0}^{\frac{\beta-1}{2}}}{K_{\frac{1+\beta}{2}}\left((1+\beta)\frac{m_{X}}{E_{0}}\right)\Gamma\left(1+\frac{\beta}{2}\right)}\,,
\end{split}
\label{eq:spectrum}
\end{equation}
where FIPs are emitted with a modified blackbody spectrum with $E_{0}$ parameter related to the FIP average energy, a FIP mass $m_{X}>2m_{e}$ and a spectral index $\beta$. In the normalization factor, $K_{\frac{1+\beta}{2}}$ is the modified Bessel function of the second kind of order $(1+\beta)/2$ and $\Gamma$ is the Euler-Gamma function. This flux is normalized such that 
\begin{equation}
    \int_{m_{X}/2}^{\infty} dE_{e}\frac{dN_{e}}{dE_{e}}=N_{e}\,,
\end{equation}
where with $N_{e}$ we denote the number of electrons, equal to the number of positrons, produced in SN explosions by FIP decays, i.e.~$N_{e}=N_{e^{+}}=N_{e^{-}}$. 
Note that the simple prescription in Eq.~\eqref{eq:spectrum} does not depend on the type of FIP model, and the examples that we mentioned can be parametrized in this model independent way.

The lepton fluxes injected by FIPs are affected by propagation effects in the Galaxy. This is taken into account by solving the diffusion equation of these particles with the {\tt DRAGON2}~\footnote{\label{note1}\url{https://github.com/cosmicrays/DRAGON2-Beta\_version}} code~\cite{Evoli:2016xgn,Evoli:2017vim}, 
with the assumption that FIPs decay close to their source, at a distance smaller than a few kpc~\cite{Calore:2021klc}. Thus, we simulate the leptonic injection as happening directly at the SN with an energy spectrum defined as in Eq.~\eqref{eq:spectrum} and following the spatial SN distribution evaluated in Ref.~\cite{Ferriere:2007yq} convolved with the Steiman-Cameron distribution~\cite{Steiman-Cameron:2010iuq} of the spiral arms (four-arm model).

The injected flux, i.e. the number of electrons/positrons per unit area, time and energy, is written as 
\begin{equation}
\frac{d\Phi_e}{dE_e}=\frac{\Gamma_{cc}}{4\pi d_{SN}^2}  \frac{dN_{e}}{dE_{e}} \, ,
    \label{eq:injection}
\end{equation}
where $\Gamma_{cc} = 2$ SN per century is the galactic rate of core-collapse SN explosions~\cite{Rozwadowska:2020nab} and $d_{SN}=10.2$~kpc is an effective length-scale that allows us to calculate the total area of emission from the SN and is roughly the average distance between galactic SN calculated from the source distribution employed in this work. This parameter is obtained by imposing that the injected number of particles per unit of energy in Eq.~\eqref{eq:spectrum} is equal to the integral of the total flux density of particles (obtained convolving Eq.~\eqref{eq:spectrum} with the spatial distribution of sources) over the volume of the Galaxy.
Regarding the propagation, we employ a spatially-constant diffusion coefficient derived from the analyses of secondary-to-primary flux ratios in Ref.~\cite{delaTorreLuque:2022vhm} and adapted in Ref.~\cite{DelaTorreLuque:2023zyd} for the spiral arm structure of the source, gas and magnetic field galactic distributions (B/C best-fit model). In particular, for the diffusion coefficient defined in Eq.~(3.3) of Ref.~\cite{delaTorreLuque:2022vhm}, the parameters are shown in Tab.~\ref{tab:params}.
We adopt the magnetic field model derived by Ref.~\cite{Pshirkov:2011um}, with a normalization of the disk, halo and turbulent magnetic field intensities set to the values found in Ref.~\cite{DiBernardo:2012zu} from the study of synchrotron radiation. The energy density distribution of the radiation fields has been taken from Ref.~\cite{Porter:2008ve}. 
A key point to assume a stationary and smooth lepton injection is the observation that electrons and positrons propagate and interact in the galactic environment on a timescale of Gyr, extremely long compared to the SN explosions rate. Therefore, we can model the lepton injection as time-independent and smoothly following the SN distribution.

\begin{table}[t!]
    \centering
    \begin{tabular}{|c|c|c|}
    \hline
Norm. Energy    &    $E_{0}$ & $4$~GeV \\
Diffusion coeff.&    $D_{0}$  & $1.02\times10^{29}$ cm$^{2}$ s$^{-1}$ \\
Diffusion index &    $\delta$&$0.49$  \\
Break energy &       $E_{b}$&$312$~GeV \\
Index break &        $\Delta\delta$&$0.20$ \\
Smooth. param. &     $s$&$0.04$\\
$\beta$ exponent &  $\eta$&      $-0.75$\\
Halo height&      $H$&$8$~kpc\\
Alfvèn velocity&  $V_{A}$&$13.4$ km s$^{-1}$\\
\hline
\end{tabular}
\caption{Main propagation parameters used in our analysis~\cite{DelaTorreLuque:2023zyd}. }
\label{tab:params}
\end{table}

Our results are robust at the level of tens of percent with respect to uncertainties in the source distribution, parameters in the diffusion coefficient and Galactic halo height, the main uncertainties come from the determination of the Alfvèn speed, which constitutes an important variable for our predictions. The Alfvèn speed is the main parameter characterizing the reacceleration of charged particles when interacting with plasma turbulences in the Galaxy~\cite{1995ApJ...441..209H, 1998ApJ...509..212S, 2019Galax...7...49C, 2022ApJ...941...65B}. Different CR analyses obtain very different values of this parameter, depending on the spallation cross sections employed or the datasets analyzed, ranging anywhere from V$_A\sim0$~km/s to V$_A\sim40$~km/s~\cite{delaTorreLuque:2022vhm, Luque:2021nxb,Luque:2021ddh}. This makes our predicted signals uncertain by orders of magnitude around the GeV scale. We have shown this in Fig.~8 of our companion work~\cite{DelaTorreLuque:2023huu}. 
Something similar was observed in Ref.~\cite{2017PhRvL.119b1103B} for the case of low mass WIMPs annihilating or decaying via electrophilic channels. The effect of reacceleration in the electron/positron signals produced by these particles was shown to cause differences of several orders of magnitude in the extracted limits on the annihilation rate for different reacceleration setups.
We remark that the limit of $V_A=0$ km/s is actually nonphysically pessimistic limit, because the observed CR diffusion implies that plasma waves are interacting with the CRs and exchanging energy, thus producing non-vanishing reacceleration. We do not include values greater than $V_A=40$ km/s because the recent detailed CR analyses do not predict values much larger than these ones. 

\emph{XMM-Newton constraints on the exotic lepton injection.} During their propagation in the Milky Way the electrons/positrons injected by FIPs interact with the interstellar gas, the interstellar radiation fields (ISRFs) 
and the Galactic magnetic field.
The IC emission produced from the boost of the low energy ISRF photons produce hard X-rays, at keV scale, and low-energy $\gamma$-rays.
For example, a typical IR photon with an energy around $0.1$~eV would gain an energy factor of $(E_e/m_e)^2 \sim 10^{4}$ when interacting with an electron of $E\sim50$~MeV, acquiring an energy around the keV-scale. At slightly higher energies, the IR field dominates the emission, and then the optical and UV fields (with smaller energy density though) take over.
As we show in Fig.~4 of our companion work~\cite{DelaTorreLuque:2023huu}, one can benefit from MeV $\gamma$-ray observations as well as X-ray observations in order to probe these signals and the injected electron/positron population, setting constraints on the injected number of electrons $N_e$.

In this study, we model the secondary diffuse emission due to the interaction of the injected leptons using the {\tt HERMES} code~\cite{Dundovic:2021ryb}, as in Ref.~\cite{DelaTorreLuque:2023huu}.
The effect of reacceleration in the electron spectrum causes a significant increase in the X-ray emission from the IC radiation of ISRF photons. In this work, we show how reacceleration boosts these signals and we derive constraints on the injection of electrons/positrons using data from the MOS detector~\cite{Turner:2000jy} on board the XMM-Newton telescope~\cite{XMM:2001haf}.

X-ray observations allow us to constrain these signals, benefiting from lower systematic uncertainties. In particular, we use the data from the MOS detector in the energy range from $1$~keV to above $10$~keV, which was provided by Ref.~\cite{Foster:2021ngm} divided in galactocentric rings around the Galactic center that are $6^{\circ}$ wide. 
In Fig.~\ref{fig:XMM-Masses} we compare the calculated FIP signals, for FIP masses between $2$ and $20$~MeV, with MOS data at the most constraining ring (ring 3) and the outermost ring (ring 30). The X-ray flux produced from FIPs of such masses differ by no more than $\sim10\%$, whereas the energy trend of the emission is roughly the same. 
We derive bounds on $N_e$ at $95\%$ C.L. from a $\chi^2$ analysis of the data. Precisely, we impose the $2\sigma$ bound on the FIP parameter space when
\begin{equation}
    \sum_i \left( \frac{\textrm{Max}\left[\phi_{\textrm{X} i}(m_{X}) - \phi_{i}, 0 \right]}{\sigma_i}\right)^2 =4
\end{equation}
where $i$ labels the data point, $\phi_{i}$ is the observed flux and $\sigma_i$ the standard deviation on its measurements.
Following Refs.~\cite{Foster:2021ngm, Cirelli:2023tnx}, we conservatively consider only datapoints in the range from $2.5$ to $8$~keV. Outside this energy range, observations are affected by instrumental noise. In Tab.~\ref{tab:XGbounds} we show the limits obtained for a $m_X = 10$~MeV FIP for different rings and Alfvèn velocities. On top of this, combining data and predictions for all the rings allow us to set a constraint that a factor $\sim2$ stronger than the one obtained from the ring 3 alone. 

\begin{figure}[t!]
\includegraphics[width=\linewidth]{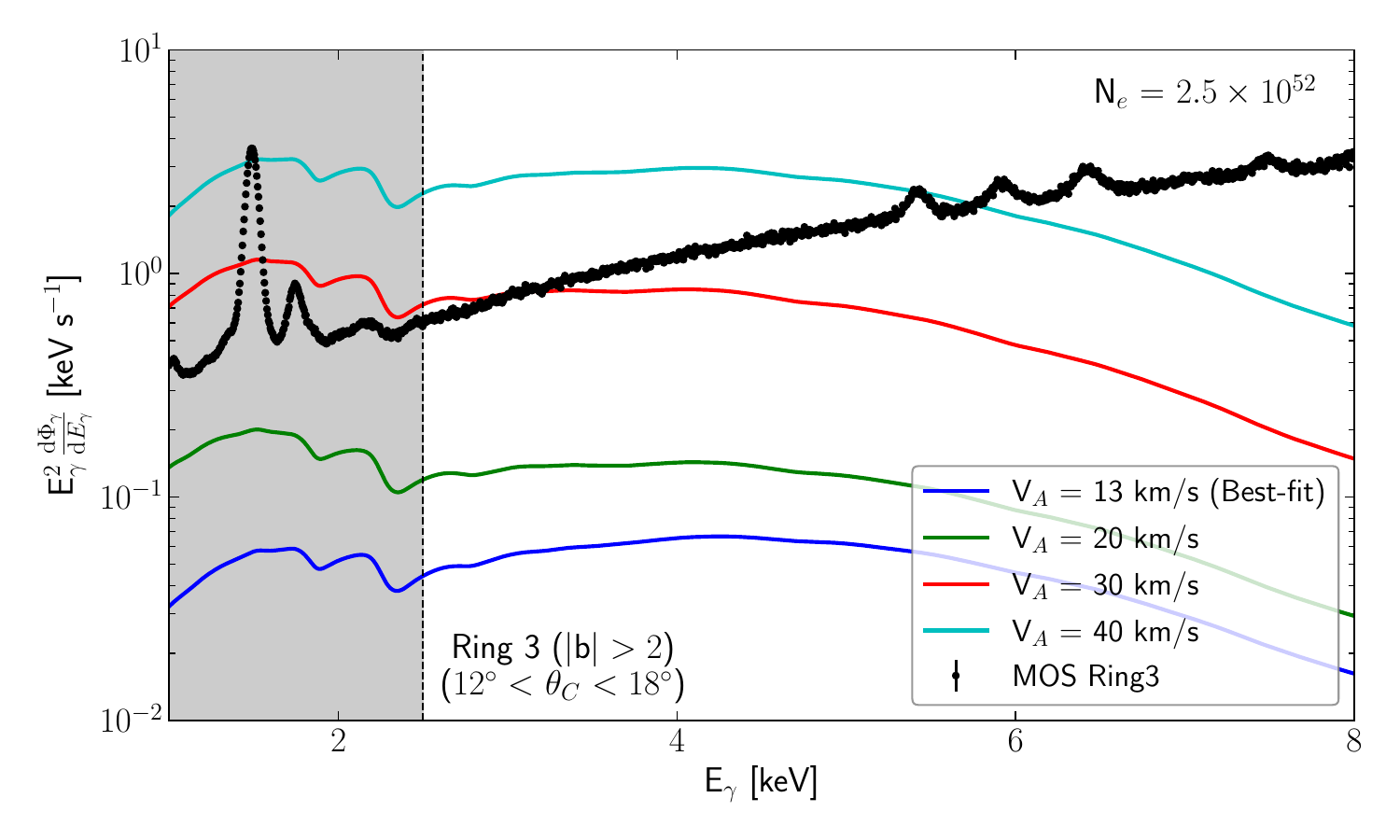}
\caption{Similar to what is shown in Fig.~\ref{fig:XMM-Masses} for different levels of reacceleration, corresponding to Alfvèn speeds of $13$, $20$, $30$ and $40$~km/s, for a FIP mass $m_{X}=10$~MeV and $N_e=2.5 \times 10^{52}$. These limits are compared to the data from ring 3 of the MOS detector.}
\label{fig:XMMvsVa}
\end{figure}

In Fig.~\ref{fig:XMMvsVa}, we show the predicted X-ray signal generated from the electrons/positrons produced by FIPs for different reacceleration levels, corresponding to Alfvèn speeds of $13$ (our reference value), $20$, $30$ and $40$~km/s, for an injection of $N_e=2.5 \times 10^{52}$. 
\begin{table}[t!]
\resizebox*{0.8\columnwidth}{0.15\textheight}{
\begin{tabular}{|c|c|}
\hline
 &\cellcolor{blue!25}\textbf{$N_{e}(\times10^{53})$ $95\%$ C.L.}  \\ 
\hline
{Best-fit V$_A$ (ring 1)}  & $37$  \\
\hline
{Best-fit V$_A$ (ring 3)} & $4.7$  \\       
\hline
{Best-fit V$_A$ (ring 4)} & $6.1$  \\       
\hline
{Best-fit V$_A$ (ring 30)} & $1015$  \\   
\hline
{Best-fit V$_A$ (Combined)} & $2.8$  \\       
\hline
{V$_A = 20$~km/s (Comb.)} & $1.1$  \\       
\hline
{V$_A = 30$~km/s (Comb.)} & $0.18$  \\       
\hline
{V$_A = 40$~km/s (Comb.)} & $0.056$  \\       
\hline
\end{tabular}
}
\caption{Limits on $N_e$ at the $95\%$ C.L. derived from MOS data between $2.5$~keV and $8$~keV. Results for various rings and Alfvèn velocities are shown.}
\label{tab:XGbounds}
\end{table}
This, in addition to the fact that MOS measurements and its uncertainties are much more robust than those evaluated for the $511$~keV line makes the MOS instrument exceptionally valuable for probing FIP physics.

Finally, we may comment again on Fig.~\ref{fig:Limits_Arrow}, where we compare the limits obtained for the number of electrons/positrons injected by FIPs with masses ranging from $\sim1$~MeV to $\sim20$~MeV from different observables. We added hatched regions indicating the uncertainties in our predictions (see Ref.~\cite{DelaTorreLuque:2023huu} for a complete description of the Voyager-1 and SPI limits). We also indicate the limit obtained for different values of the Alfvèn speed, ranging from $\sim 10$~km/s to $40$~km/s.
Notably, the data from the MOS detector (combining all the rings) leads to constraints on $N_e$ ranging from $\sim3\times 10^{53}$ to $\sim 5 \times 10^{53}$, which are compatible and even better, in the case of high reacceleration, to those derived from the longitude profile of the $511$~keV line. With the latter being less robust than MOS constraints due to sizeable systematic uncertainties in the extraction of the data, as discussed in more detail in Ref.~\cite{DelaTorreLuque:2023huu}.

\emph{XMM-Newton constraints on FIP properties.}
In Ref.~\cite{DelaTorreLuque:2023huu} we derived constraints on specific FIP models, corresponding to our bounds on $N_e$. We compute the $2\sigma$ upper limits by a $\chi^2$ analysis of the X-ray data, as in Ref.~\cite{DelaTorreLuque:2023huu}.
Here, we briefly discuss how we expect the constraints on FIPs to improve with the proposed bounds. 

An useful simplification is to consider constraints in the weak coupling regime, i.e. where the positron production per SN grows as $g^{2}$ with $g$ FIP-electron coupling. Thus, we neglect the strong coupling regime, with an exponential sensitivity to $g$ since FIPs starts to decay inside the SN and part of the produced electron-positron pairs remains trapped inside the star. Under this assumption, the number of electrons/positrons per SN is proportional to some power of the coupling, $N_{e}\sim g^{\alpha}$, where $\alpha$ might be different from $2$ in order to schematically take into account the effect of type Ib/c SNe. Precisely, in this discussion we will consider sterile neutrinos mixed with muon and tau neutrinos, and DPs, all of them with a mass $m_{X}=10$~MeV. In these cases, $g=|U_{s\mu}|$ (sterile-muon neutrino mixing parameter) and $\alpha=1.64$, $g=|U_{s\tau}|$ and $\alpha=1.52$,  $g=\epsilon$ (kinetic mixing parameter with the photon) and $\alpha=2$, respectively~\cite{DelaTorreLuque:2023huu,Carenza:2023old}. 
In the case of ALPs we can only determine a bound in the strongly coupled regime because of our assumption of FIPs decaying close to the production point, that is violated in the weak coupling regime.
Thus, we do not discuss ALP constraints here.

\begin{figure}[t!]
\includegraphics[width=\linewidth]{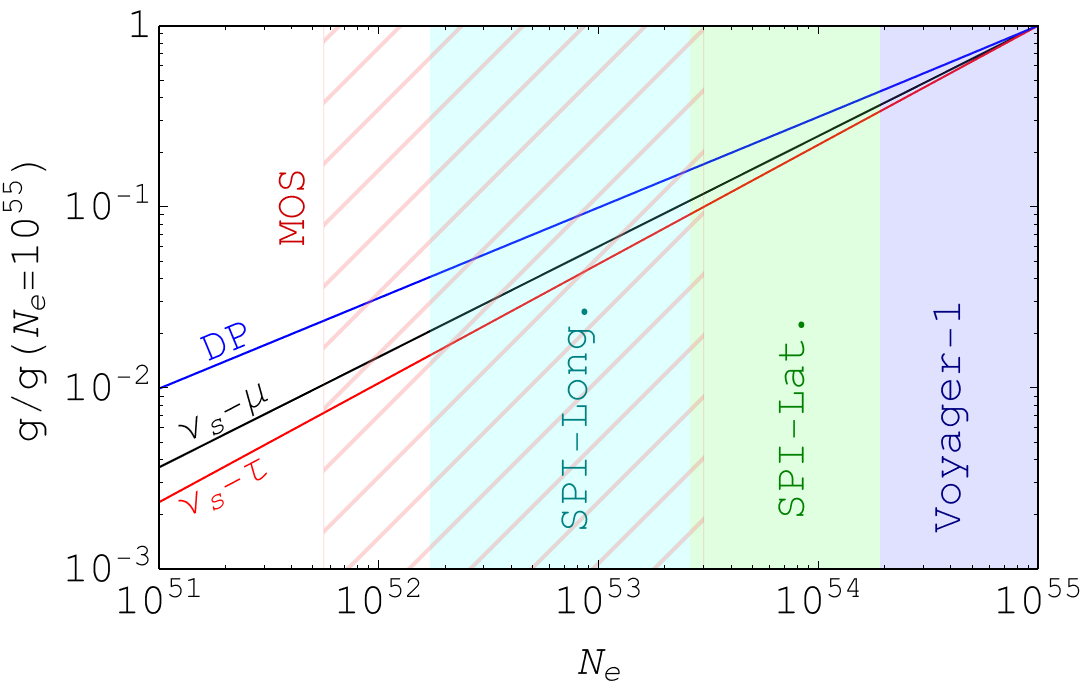}
\caption{Improvement of the bounds on $g$, for various FIP models discussed in the text, as a function of the maximal $N_{e}$ allowed by observations. The 511 keV line bounds, for the latitudinal and longitudinal distributions are shown in green and cyan, respectively. The Voyager-1 bound is shown in blue and the red hatched region refers to the uncertainty band on the maximal $N_{e}$ associated with MOS observations.}
\label{fig:boundsFIP}
\end{figure}

Now it is easy to understand how the limiting values of $N_{e}$ can be reflected as  constraints on the FIP-electron couplings: $g\sim N_{e}^{1/\alpha}$. In Fig.~\ref{fig:boundsFIP} we show the improvements of the bound on $g$ as a function of $N_{e}$, with respect to a nominal bound placed with $N_{e}=10^{55}$, for sterile neutrinos (black and red lines) and DPs (blue line). These lines for various FIP models are fixed by the $N_e$-coupling relation mentioned above for FIP masses $\lesssim20$ MeV considered in this work, over which we can now superimpose various FIP bounds on $N_e$. In Fig.~\ref{fig:boundsFIP} we highlight as colored regions, the bounds discussed in this work with the same color code of  Fig.~\ref{fig:Limits_Arrow}, showing only the most stringent constraints related to the 511 keV line. The red hatched region is the uncertainty band of the MOS constraint that, under optimistic assumptions, improves the leading one by a factor $\sim3$ on $N_{e}$. In the considered models, bounds on the FIP couplings improve up to a factor $\sim2$ thanks to the MOS analysis.
We note the robustness of the derived X-ray bound, given the quality of the MOS data. In comparison, the bound derived from the $511$~keV line emission is plagued by systematic uncertainties in the data, and, even more importantly, the current evaluation of the $511$~keV signal neglects important ingredients, such as a proper positron propagation treatment, energy losses, or in-flight annihilation, that would modify the limit significantly.\\
We mention also that the fireball formation due to FIP decay into electron-positron pairs~\cite{Diamond:2023scc,Diamond:2023cto} is unlikely to happen for the majority of SNe contributing to the signal considered in this work i.e. from red supergiant progenitor~\cite{DelaTorreLuque:2023huu}. Thus, we omit this possibility here.

\emph{Conclusions.} In conclusion, we have explored for the first time, the capability of the hard X-ray observations from the XMM-Newton experiment to set constraints on the production of electrons/positrons from the decay of FIPs. In addition, we evaluated the impact of reacceleration on the electromagnetic signals generated from FIPs. We find that observations from the XMM-Newton mission allow us to set one of the strongest limits on the production of electrons and positrons from FIPs in the Galaxy, possibly even more stringent than those derived from the $511$~keV emission. This study opens the path for future searches of particles beyond Standard Model coupled with electrons and positrons in astrophysical X-ray spectra, showing a remarkable complementary with cosmological bounds. Finally, it is worth remarking the potential of the X-ray band: current data at even lower energies (such as those from Chandra), could allow us to improve these bounds by more than an order of magnitude.

\acknowledgements
\emph{Acknowledgements.}
SB would like to thank Jordan Koechler and Marco Cirelli for useful discussions regarding XMM-Newton data. This article is based upon work from COST Action COSMIC WISPers CA21106, supported by COST (European Cooperation in Science and Technology).
SB is supported by funding from the European Union’s Horizon 2020 research and innovation programme under grant agreement No.~101002846 (ERC CoG ``CosmoChart'') as well as support from the Initiative Physique des Infinis (IPI), a research training program of the Idex SUPER at Sorbonne Universit\'{e}. The work of PC is supported by the European Research Council under Grant No.~742104 and by the Swedish Research Council (VR) under grants  2018-03641 and 2019-02337. 
PDL is supported by the European Research Council under grant 742104 and the Swedish National Space Agency under contract 117/19.
This project used computing resources from the Swedish National Infrastructure for Computing (SNIC) projects 2021/3-42, 2021/6-326, 2021-1-24 and 2022/3-27 partially funded by the Swedish Research Council through grant no. 2018-05973.


\bibliographystyle{apsrev4-1}
\bibliography{references.bib}

\end{document}